\documentclass[a4paper, amsfonts, amssymb, amsmath, reprint, showkeys, nofootinbib, twoside]{revtex4-1}
\usepackage[english]{babel}
\usepackage{amsmath}
\usepackage{mathtools}
\usepackage{amssymb}
\usepackage{physics}
\usepackage{booktabs}
\usepackage{array}
\usepackage{float}
\usepackage[utf8]{inputenc}
\usepackage[colorinlistoftodos, color=green!40, prependcaption]{todonotes}
\usepackage[pdftex, pdftitle={Article}, pdfauthor={Author}]{hyperref} 
\begin{document}

\title{Quantum Random Access Code in Noisy Channels}

\author{Rafael A. da Silva}
    \email[Correspondence email address: ]{rafael.a@ufabc.edu.br}
     \affiliation{Itaú Quantum Technologies}
     \affiliation{Universidade Federal do ABC, CCNH, Santo André, SP, Brazil}
    \author{Breno Marques}
    \email[Correspondence email address: ]{breno.marques@ufabc.edu.br}
    \affiliation{Universidade Federal do ABC, CCNH, Santo André, SP, Brazil}

\date{\today} 

\begin{abstract}
Random access code (RAC) communication protocol particularly useful when the communication between parties is restricted. In this work we built upon works that have previously proven  quantum random access code (QRAC), in the absence of noise, to be more advantageous than classical random access code (CRAC), investigate the effects of noisy channel on QRAC performance and how the losses can be mitigated by using the see-saw method optimized by semi-definite programming when the noisy channel is known.


\end{abstract}

\keywords{Quantum communication, noisy channel, quantum random access code.}

\maketitle
\section{Introduction}
In quantum communication one uses quantum resources such as superposition and entanglement to enhance information transmission beyond classical limitations \cite{Gal}. An example of this is the quantum random access code (QRAC), first introduced by S. J. Wiesner in 1983 \cite{Wiesner2} and rediscovered by A. Ambaini \textit{et al.} \cite{Andris}, in which the use of quantum strategies for encoding and decoding Alice's messages improves Bob's probability of correctly accessing the information he interested in when compared to classical strategies.
In general, a QRAC, involves a party, Alice, who has an $n$-dit long string which she must encode in $m<n$ qudits and send to Bob, who is only interested in a subset of the string. He must be able to retrieve this information with average probability of success $P_d>P_{chance}$, where $P_{chance}$ represent his average probability if he randomly guessed the information. This family of QRACs can be symbolically represented as $n^{(d)} \xrightarrow{P_d} m$. Armin Tavakoli \textit{et al.} showed in \cite{Breno} that, in the noiseless regime, the $2^{(d)} \xrightarrow{P_d} 1$ QRAC outperforms its classical counterpart (the CRAC)  for any dimension $d$ in terms of  average probability of success.\\
\indent However, in the implementation quantum communication protocols, in both experimental and application contexts, it is unlikely that one will find  a system that operates so close to ideality that one may completely neglect the influence of quantum noise on said system and, consequently, on the protocol itself. Therefore, in those contexts it is fundamental to account for sources of noise for determining possible limitations on the implementation the protocol or even its viability. Previous works have addressed the issue of quantum communication through noisy channel for protocols such as QKD \cite{QKD}, quantum steganography \cite{steno}, quantum teleportation \cite{alejandro, fortes}. However, the issue of the QRACs under noisy quantum channels has not been addressed, especially not for the $2^{(d)} \xrightarrow{P_d} 1$ QRAC. 

In this work, we investigate, \textit{via} simulations, how Bob's average probability of success, $P_d$, evolves in time, for a given dimension $d$, when the communication happens through one of the following Markovian channels:  the dit flip, d-phase flip, dephasing, depolarizing, and amplitude damping channels. The simulations show that the action of these channels can reduce the efficiency of the QRAC to the point that its classical counterpart performs better. We then attempt to mitigate this loss in performance by optimizing the the protocol using semi-definite programming (SDP), a sub-field of convex optimization that has been extensively applied for a wide range of purposes in quantum information \cite{SDPCA, SDPSW,SDPRA, SDPFE}. It is especially well suited for our problem because our figure of merit, the quantum average probability of success, depends linearly on both the encoding states (density matrices) and decoding measurement operators, which are likewise positive semi-definite and are constrained to having trace equal to one. Therefore, the task of maximizing this probability of success can be cast as an SDP problem.

In section (\ref{first}) we review the classical and quantum $2^{(d)} \xrightarrow{P_d} 1$ random access code protocol. In section (\ref{second}) we presented a basic overview of theory of open quantum system and how it relates to concept of noisy quantum channels; we also introduce the three noisy quantum channels studied in this work: depolarization, amplitude damping, and dephasing channels.  In section (\ref{third}) we applied the concepts laid out in the previous section to study the behaviour and performance of quantum RAC in those noisy channels,how the noise is mitigates changing the protocol strategy.

\section{Review of Random Access Code}\label{first}

In a RAC, Bob aims to access an arbitrary subset of information held by Alice, using an restricted communication channel. The probability of success can be optimized by the chosen communication strategy \cite{Breno}.

\subsection{Classical Random Access Code}
\begin{figure}[H]
    \centering
    \includegraphics[scale=0.3]{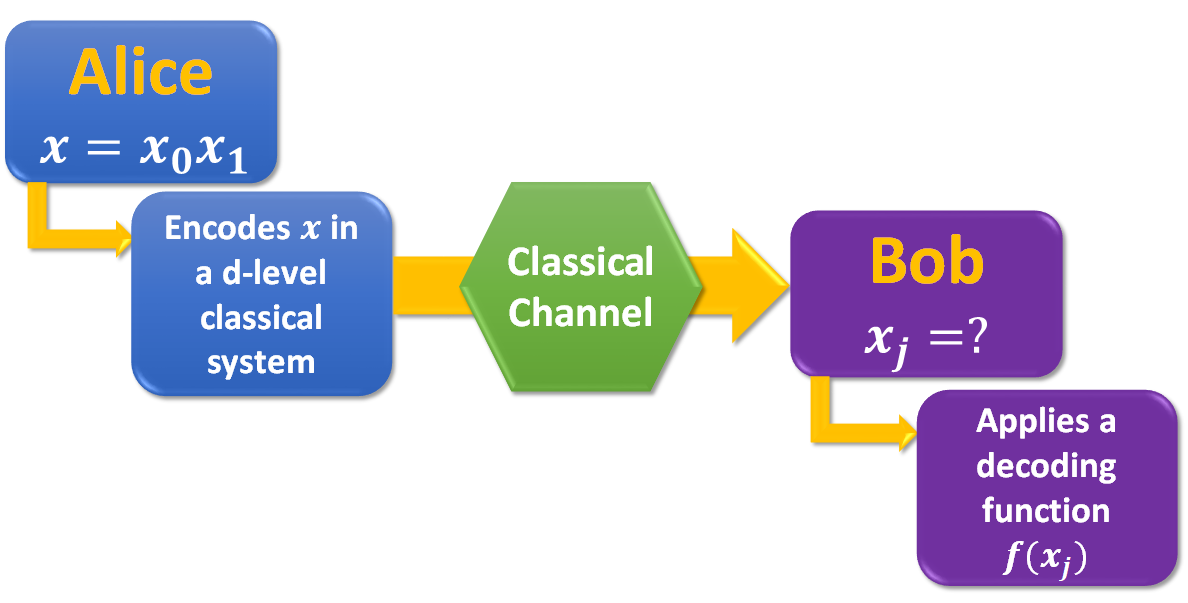}
    \caption{Schematics of a classical RAC protocol for $2^{(d)} \xrightarrow{P_d} 1$ scenario.}
    \label{fig:classico}
\end{figure}

In the classical version of the $2^{(d)} \xrightarrow{P_d} 1$ RAC, Alice encodes the string  $x=x_0x_1$ in a $d$-level classical state (see figure \ref{fig:classico}). In the best classical strategy \cite{Breno} Alice always sends the first input $x_0$. If Bob have to guess the first (second) input the probability of success will be 1 (1/d). The average probability of success will be given by:
\begin{equation}\label{eq1}
    P^C=\frac{1}{2}\left(1 + \frac{1}{d}\right).
\end{equation}


\subsection{Quantum Random Access Code}
\begin{figure}[H]
    \centering
    \includegraphics[scale=0.3]{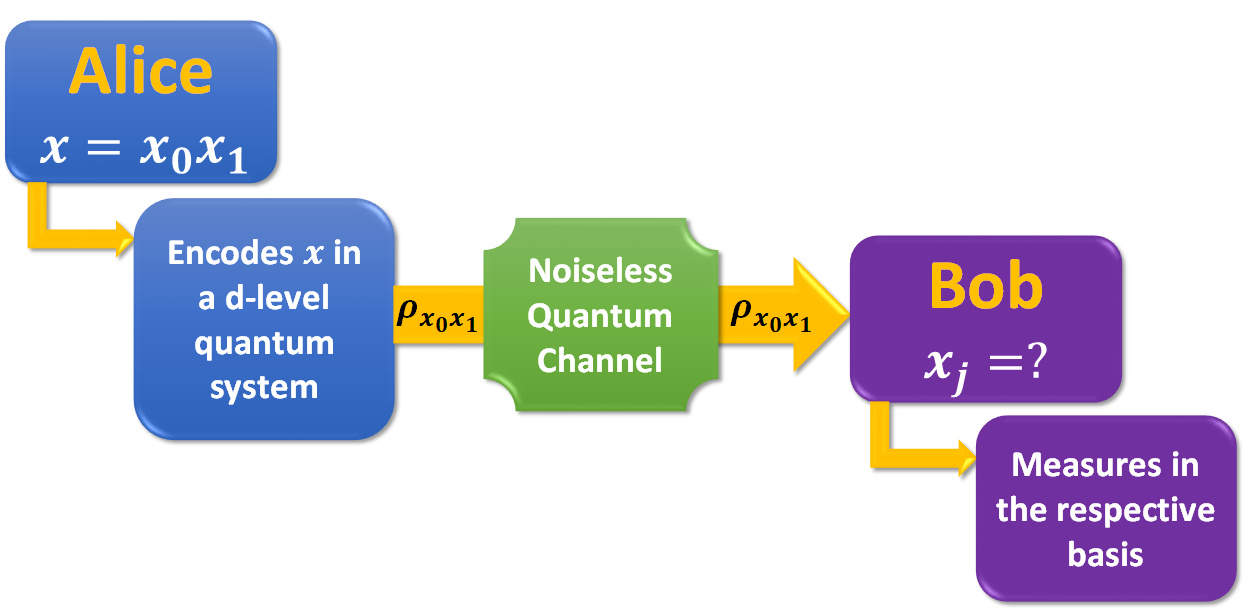}
    \caption{Schematics of a quantum RAC protocol through a noiseless communication channel for $2^{(d)} \xrightarrow{P_d} 1$ scenario.}
    \label{fig:quantico}
\end{figure}

\begin{figure}[H]
    \centering
    \includegraphics[scale=0.7]{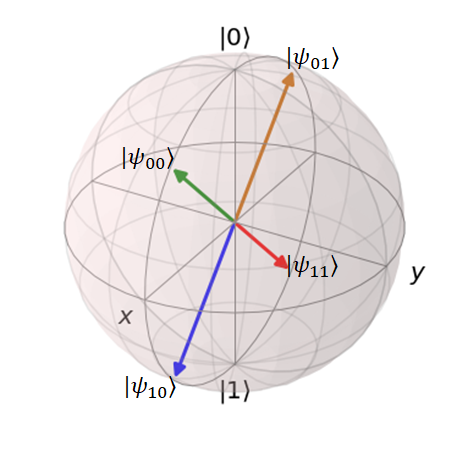}
    \caption{We present here a example for $d=2$ represented in the Bloch sphere. Notice that the projections on the $z$-axis (computational basis) and on the $x$-axis (Fourier basis) have the same modulus.}
    \label{fig:state_distribution}
\end{figure}

In the quantum version of the $2^{(d)} \xrightarrow{P_d} 1$ RAC, however, Alice encodes $x=x_0x_1$ in a single $d$-level quantum state (see figure \ref{fig:quantico}) \cite{Breno}. This state can be constructed in terms of two mutually unbiased bases (MUB). In the present work we chose the computational  $B_e=\{\ket{e_l}\}_{l=0}^{d-1}$ ($\ket{e_l}=\ket{l}$), and the Fourier basis $B_f=\{\ket{f_l}\}_{l=0}^{d-1}$ ($\ket{f_l}=(1/\sqrt{d})\sum_{n=0}^{n=d-1}\omega^{ln}\ket{n}$, where $\omega=\exp{2\pi i/d}$) for constructing the encoding states
\begin{equation}\label{eq2}
      \ket{\psi_{x_0x_1}}=\frac{1}{\sqrt{2+(2/\sqrt{d})}}\left(\ket{e_{x_0}} + \ket{f_{x_1}}\right).
\end{equation}
Whenever Bob is interested in $x_0$ ($x_1$), he performs a measurement in the basis $B_e$ ($B_f$), and the average probability of success for this strategy is given by:
\begin{equation}\label{eq3}
 \begin{split}
     \MoveEqLeft
      P^Q=\frac{1}{2d^2}\sum_{x_0=0}^{d-1}\sum_{x_1=0}^{d-1}\Tr{\rho_{x_0x_1}(M_{x_0}+M_{x_1})} \\
      &=\frac{1}{2}\left(1 + \frac{1}{\sqrt{d}}\right),
 \end{split}
\end{equation}
where $\rho_{x_0x_1}=\op{\psi_{x_0x_1}}$, $M_{x_0}=\op{e_{x_0}}$ and $M_{x_1}=\op{f_{x_1}}$.
It is easy to show that the probabilities of all outcomes are the same (equal to $P^Q$) regardless of Alice's string $x=x_0x_1$ and Bob's input guessed. This happens because the magnitudes of the projections of $\ket{\psi_{x_0x_1}}$ into $B_e$ and $B_f$ basis are all equal (see figure \ref{fig:state_distribution}, where this can be illustrated for $d=2$ using the Bloch sphere). As we will see sections \ref{second} and \ref{third}, this symmetry can be broken when the communication occurs thought noisy channels.


Comparing equations (\ref{eq1}) and (\ref{eq3}). It is straightforward to see that, in the noiseless communication channel, the probability of success is always greater for quantum case than the classical case, \textit{i.e.}, 
\begin{equation}\label{pqpc}
    P^Q/P^C>1 \quad\forall d.
\end{equation}
This ratio will be from now on the figure of merit when assessing the performance of the QRAC.

\section{Noisy quantum channels}\label{second}
In non-ideal conditions, quantum noise is a limiting, or even prohibiting, the quantum communication protocols enhancement. Understanding how a quantum noise channel can affect the protocol is essential for predicting noise-related efficiency loss and it might be possible to find strategies for mitigating said effects without enhance the protocol complexity, e.g., using quantum error correction protocols.  The resulting dynamics can be understood in the light of the theory of quantum channels \cite{Petruccione,crispin} that can be understand as a interaction between the encoding quantum system and the environment. This theory establishes that quantum channels in general, are  completely positive and trace-preserving (CPTP) maps \cite{Capacity} which can be written in the Kraus representation as 
\begin{equation}
    \mathcal{N}(\rho)=\sum_{\nu}K_{\nu}\rho K_{\nu}^{\dagger},
\end{equation}
where $K_{\nu}$ are the so-called Kraus operators, which satisfy $\sum_{\nu}K_{\nu}K_{\nu}^{\dagger}=I$. The quantum channels that will be considered in this work are presented in the next subsections and, as an example, the possible accessible states after the noise channel will be shown in a Block sphere representation for $d=2$.

\subsection{Dit-flip Channel}
\begin{figure}[H]
    \centering
    \includegraphics[scale=2.1]{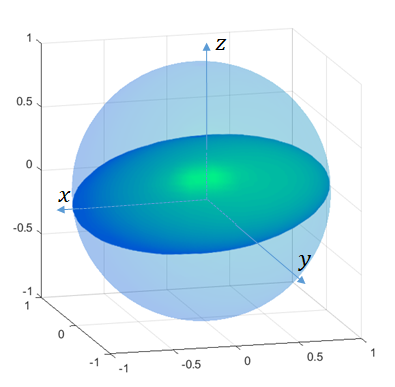}
    \caption{Representation on the Bloch sphere of the dit-flip channel for the qubit. This map conserves the probabilities of measurements in the Fourier basis but decreases the probabilities for measurements in the computational basis.}
    \label{fig:dit}
\end{figure}

This channel is the d-dimensional generalization of the bit-flip channel whose action flips a qudit $\ket{\mu}$ (figure~\ref{fig:dit}), with equal probability, to one of the states $\ket{\mu\oplus 1}$, $\ket{\mu\oplus 2}$, ..., $\ket{\mu\oplus d-1}$. Since the dit flip channel belongs to the family of discrete Weyl's channels (DWC)  \cite{alejandro, Rehman}, it is possible to use the Weyl's operators
\begin{equation}
    W_{\nu\mu}=\sum_{k=0}^{d-1}\omega^{k\nu}\op{k}{k\oplus \mu},\quad \omega=\exp{2\pi i/d},
\end{equation}
to write its set of Kraus operators as 
\begin{equation}
        K_{\nu}=
        \begin{cases}
            \sqrt{1-p}W_{0\nu}, &\nu=0 \\
            \sqrt{\frac{p}{d-1}}W_{0\nu}, &1\leq \nu \leq d-1.
        \end{cases}
    \end{equation}

\subsection{D-phase-flip Channel}
\begin{figure}[H]
    \centering
    \includegraphics[scale=0.81]{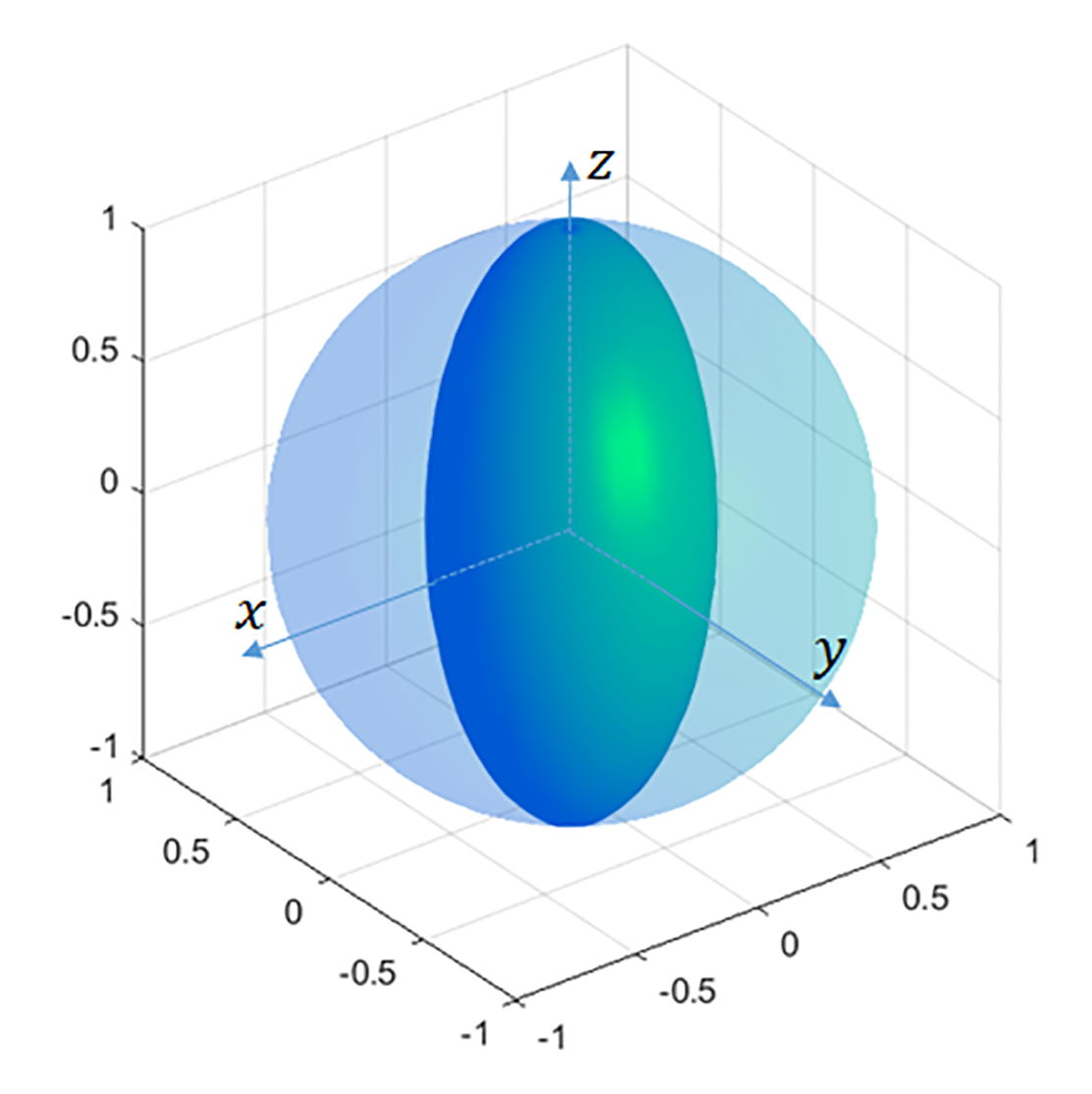}
    \caption{Representation on the Bloch sphere of the phase flip channel for the qubit. This map conserves the probabilities of measurements in the computational basis but decreases the probabilities for measurements in the Fourier basis.}
    \label{fig:phase}
\end{figure}

The d-phase-flip channel is a generalization of the qubit phase-flip channel (figure~\ref{fig:phase}). It acts on a qudit $\ket{\mu}$ flipping its phase, with equal probability, in one of the following ways $\omega\ket{\mu}$, $\omega^2\ket{\mu}$, ..., $\omega^{d-1}\ket{\mu}$, with $\omega=\exp{2\pi i/d}$. The Kraus operators for this channel are given as
\begin{equation}
        K_{\mu}=
        \begin{cases}
            \sqrt{1-p}W_{\mu 0}, &\mu=0 \\
            \sqrt{\frac{p}{d-1}}W_{\mu 0}, &1\leq \mu \leq d-1.
        \end{cases}
    \end{equation}
    
\subsection{Depolarizing Channel}

\begin{figure}[H]
    \centering
    \includegraphics[scale=0.96]{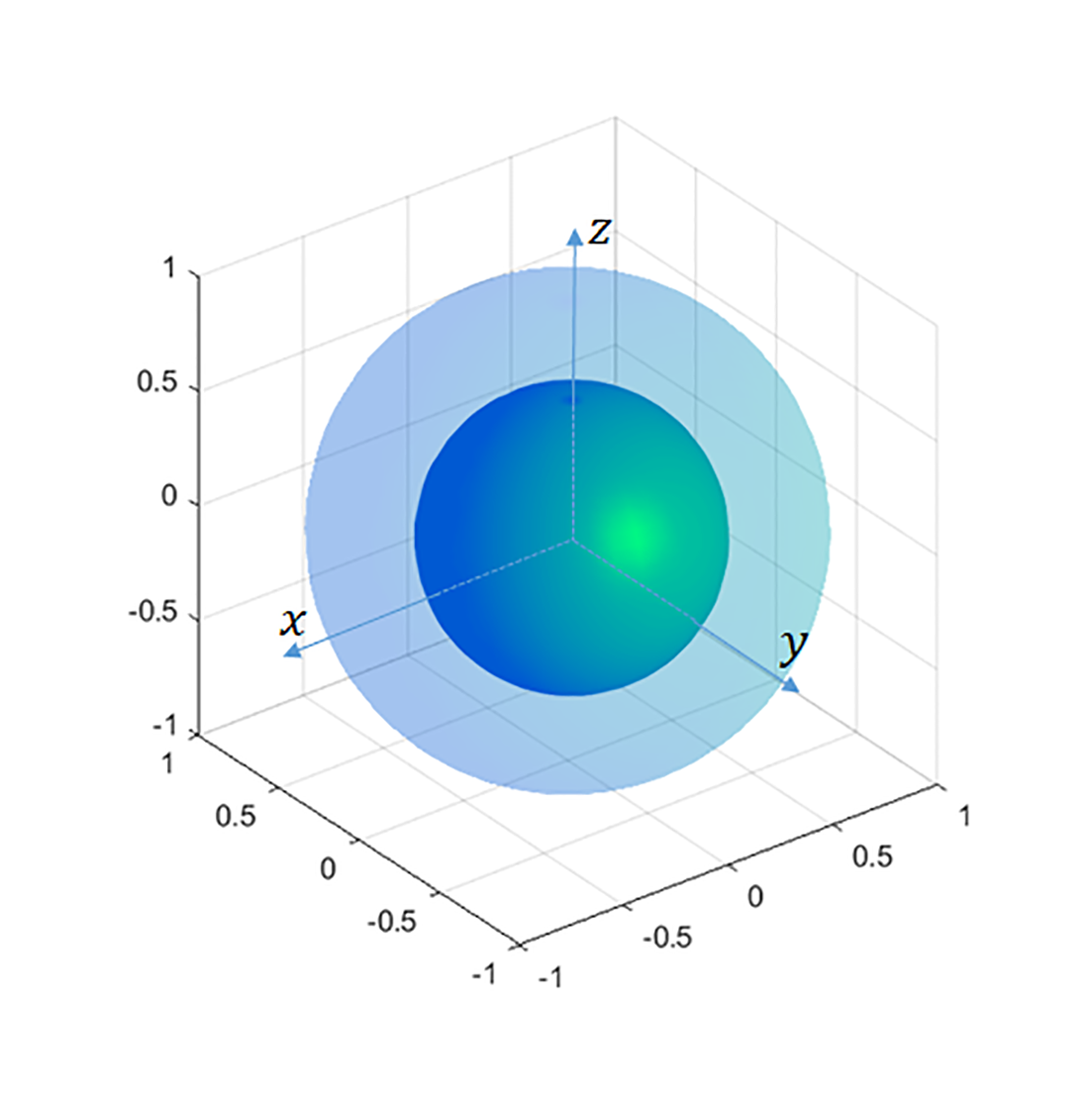}
    \caption{Representation on the Bloch Sphere of a depolarizing channel for qubit. At $t=t_0$ the states- which are pure - lie on the surface of the sphere (represented by the dotted lines). As time passes and the states become mixed under the influence of noise, they now lie inside the old Bloch sphere as if it had shrunk. At a long enough time any initial state will evolve to a completely mix state represented by a single point at the center of the old Bloch sphere.}
    \label{fig:depo}
\end{figure}

The depolarizing channel is important in experimental contexts, where it is used to analyse experimental setups in which the quantum state may be lost or when working with non-ideal detectors. For qudits ($d$-level quantum systems), this channel can be described as follows: $\rho_S$ has probability $p$ of being replaced with a completely mixed state, $I/d$, otherwise it remains unchanged . The corresponding map is
\begin{equation}
    \rho_S(t)=\frac{pI}{d}+(1-p)\rho_S(0),
\end{equation}
where $p=1-e^{-\Gamma t}$, and $\Gamma$ is the system-environment coupling constant. One interesting aspect of this channel is its symmetry (which can be visualized in figure~\ref{fig:depo} for $d=2$). As a consequence,  the probabilities of success for measurements in either $B_e$ or $B_f$ deplete in time at the same rate.

\subsection{Dephasing Channel}
The dephasing channel describes a decoherence process in which quantum information is lost without loss of energy. The evolution of a qudit under this channel can be described by
 \begin{equation}\label{masterdeph}
    \dot{\rho}_S = \Gamma\left[2a^{\dagger}a\rho_S a^{\dagger}a-\acomm{(a^{\dagger}a)^2}{\rho_S}\right],
\end{equation}
where $\Gamma$ is the dephasing system-environment coupling constant, and $a$ and $a^{\dagger}$ are the annihilation and creation operators, respectively.
By solving (\ref{masterdeph}) one will find that the elements of the initial density matrix $\rho_S(0)$ will evolve as,
\begin{equation}\label{deph_master}
   \mel{n}{\rho_S(t)}{m} =(1-p)^{(n-m)^2}\mel{n}{\rho_S(0)}{m},
\end{equation}
where $(1-p)=e^{-\Gamma t}$.

\begin{figure}[H]
    \centering
    \includegraphics[scale=2.1]{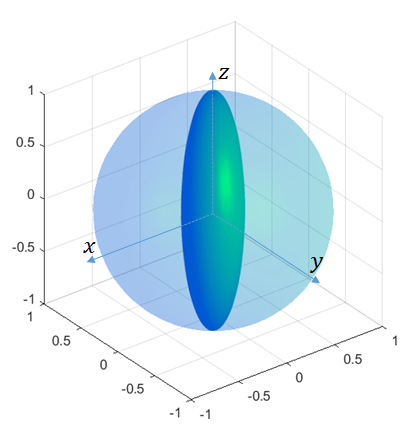}
    \caption{Representation on the Bloch sphere of the dephasing channel for the qubit. This map shrinks the Bloch sphere in both $x$ and $y$ direction while the poles remains unchanged. Again, this means that the probabilities of measurements in the computational basis are constant in time, but vary for measurements in the Fourier basis.}
    \label{fig:deph}
\end{figure}

The evolution of a qudit under this channel happens in such a manner that the probabilities for measurements in the basis $B_e$ remain constant in time while for measurements in the basis $B_f$ they decrease (see figure (\ref{fig:deph}) for $d=2$).

\subsection{Amplitude Damping}
\begin{figure}[H]
    \centering
    \includegraphics[scale=0.95]{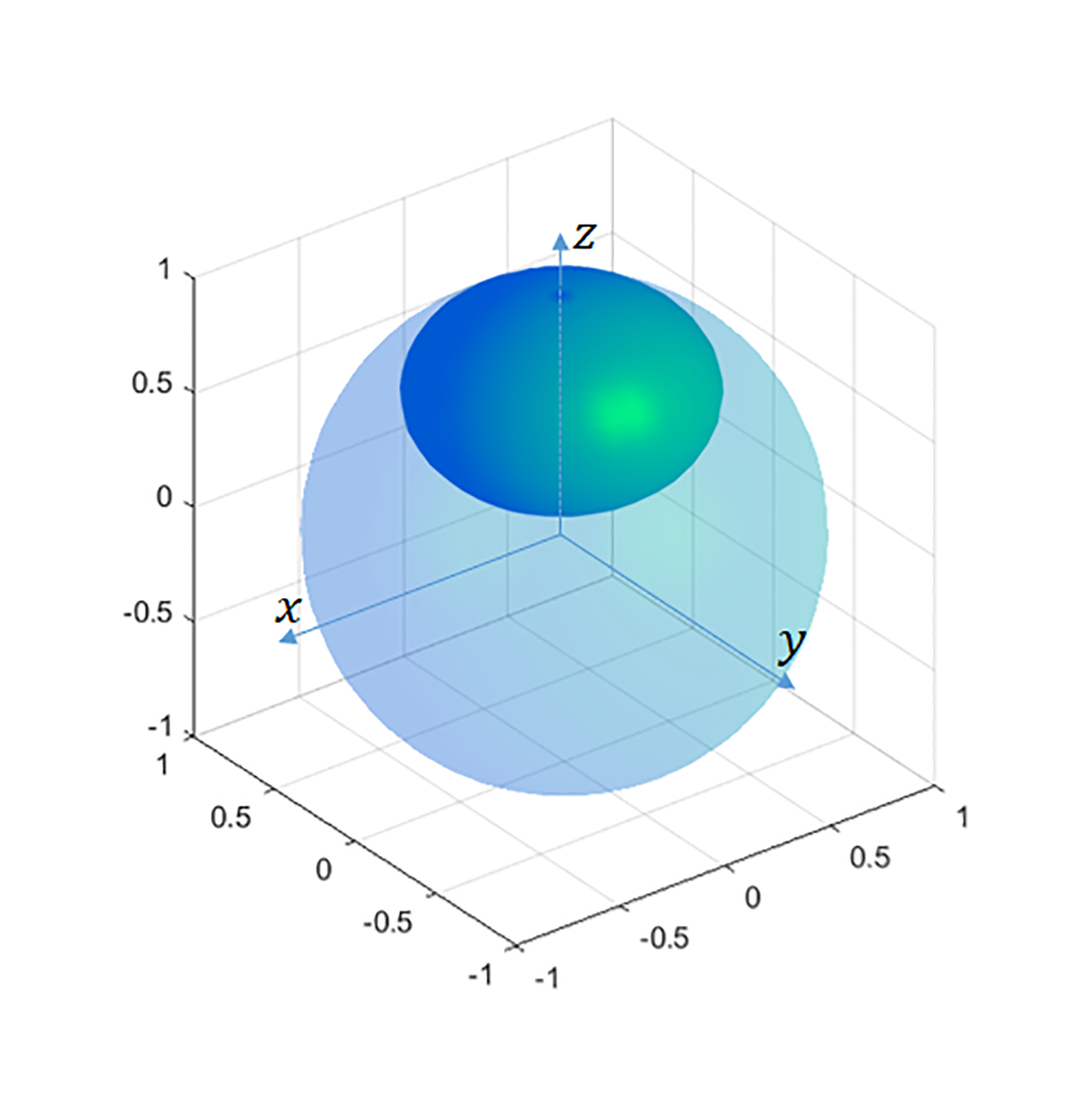}
    \caption{Representation on the Bloch sphere of the amplitude-damping channel for a qubit. This map shrinks the Bloch sphere in the two directions of the equatorial plane, and also moves the center of resultant ellipsoid towards the north pole, \textit{i.e.}, the fundamental state for spontaneous emission.}
    \label{fig:amp}
\end{figure}

The amplitude damping channel models loss of energy of the system to its environment and describes, for instance, the phenomenon of spontaneous emission \cite{Davi}. In this case, it drives the system to its fundamental state $\ket{0}$ (see figure (\ref{fig:amp}) for $d=2$), and, as consequence, the probabilities for measurements in both bases $B_e$ and $B_f$ will decrease in time. This dynamics can be expressed \textit{via} the Kraus operators
\begin{equation}
        K_{\nu}=
        \begin{cases}
            \op{\nu}+\sqrt{1-p}\sum_{k=1}^{d-1}\op{k},\quad \text{for} &\nu=0 \\
            \sqrt{p}\op{0}{\nu},\quad \text{for}\quad 1\leq \nu \leq d-1.
        \end{cases}
    \end{equation}

\section{The QRAC under noisy channels}\label{third}
In the noiseless regime, the $2^{(d)} \xrightarrow{P_d} 1$ QRAC always outperforms its classical counterpart (section \ref{first}). As expected, we will show that this is not the case when the communication is performed over noisy quantum channels. We investigate the influence of quantum noise by considering a three-step communication process composed of encoding (state preparation), transmission, and decoding (measurement). We considered noise disturbance only in transmission step.

\subsection{The non-optimized scenario}
\begin{figure}[H]
    \centering
    \includegraphics[scale=0.3]{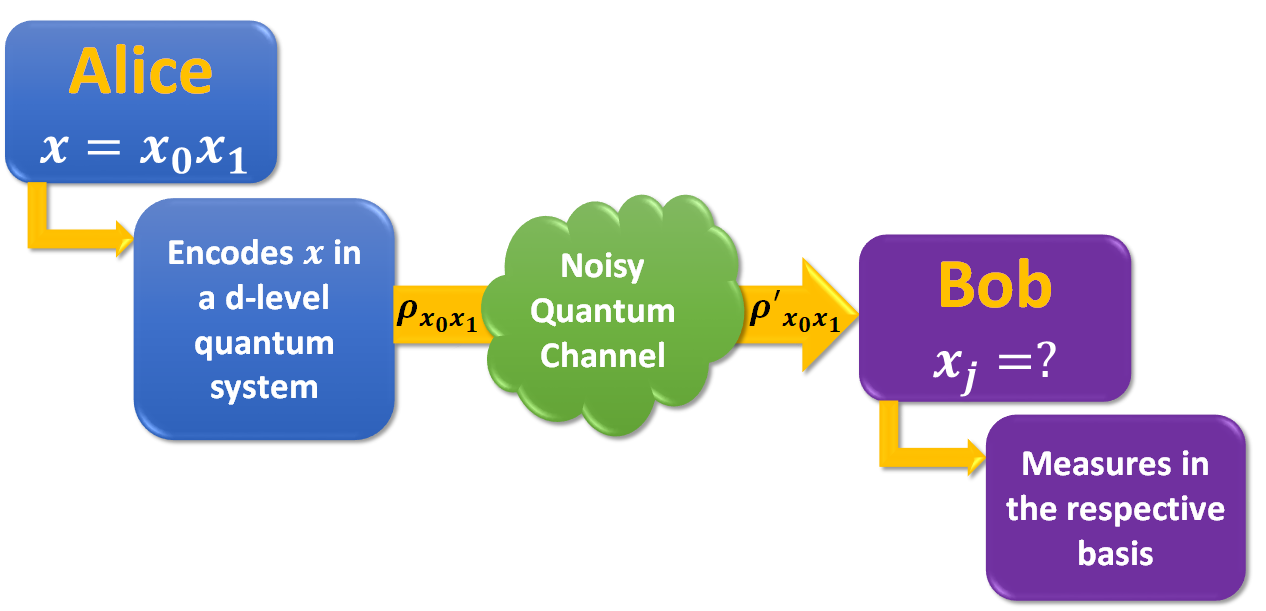}
    \caption{Schematics of a quantum RAC protocol through a noisy channel.}
    \label{fig:quantico_noise}
\end{figure}

Suppose that Alice prepares her encoding state $\rho_{x_0x_1}$ and the quantum system is sent through a noisy channel. Bob performs a measurement in either basis $B_e$ or $B_f$ as before (see figure (\ref{fig:quantico_noise}) for an illustration). Thus, his average probability of success is given by
\begin{equation}
    P^Q(t)=\frac{1}{2d^2}\sum_{x_0=0}^{d-1}\sum_{x_1=0}^{d-1}\Tr{\rho_{x_0x_1}(t)(M_{x_0}+M_{x_1})},
    \label{pqt}
\end{equation}
where $\rho_{x_0x_1}(t)$ is the quantum state received and $P^Q(t)$ can be used to evaluate how the QRAC performance changes as a function of time in comparison to the CRAC that is given by the ratio $P^Q(t)/P^C$. An important parameter is the time when QRAC losses the advantage over CRAC, $P^Q(t)/P^C=1$, that is shown in table~\ref{tab:1} for different $2^{(d)} \xrightarrow{P_d} 1$ QRAC scenario and noise channels. The rate at which this happens depends strongly on the dimension and the noise channel. Although the performance presented above is seemingly inevitable, we must entertain the possibility that our strategy of encoding and decoding using the computational and Fourier bases, albeit optimal for the noiseless case, may not be adequate using a noise channel. A  better strategy might be found when a flexible encoding and decoding are fine-tuned for each of the noisy channels to minimize the performance losses. In the following section we discuss how we applied this strategy.


 \begin{table}[h!]
    \caption{Time span (in units $\Gamma$) when $P^Q(t)P^C=1$ for different quantum system dimensions $d$ and noise channels.} 
    \centering 

    \begin{tabular}{| c | c | c | c | c | c | c |} 
    \hline\hline 
    Channel & $d=2$ & $d=3$ & $d=4$ & $d=5$ & $d=6$ & $d=7$ \\ 
    [0.5ex] 
    \hline 
    Dit flip & 0.35 & 0.44 & 0.47 & 0.48 & 0.47 & 0.46\\
    D-phase flip & 0.34 & 0.45 & 0.46 & 0.48 & 0.46 & 0.46 \\
    Dephasing & 0.88 & 0.48 & 0.28 & 0.18 & 0.12 & 0.09 \\
    Amplitude damping & 0.47 & 0.32 & 0.24 & 0.18 & 0.15 & 0.12 \\
    Depolarizing & 0.35 & 0.31 & 0.29 & 0.27 & 0.25 & 0.24 \\
    \hline 
    \end{tabular}
    \label{tab:1}
    \end{table}
    

\subsection{The optimized case}

\begin{figure}[H]
    \centering
    \includegraphics[scale=0.42]{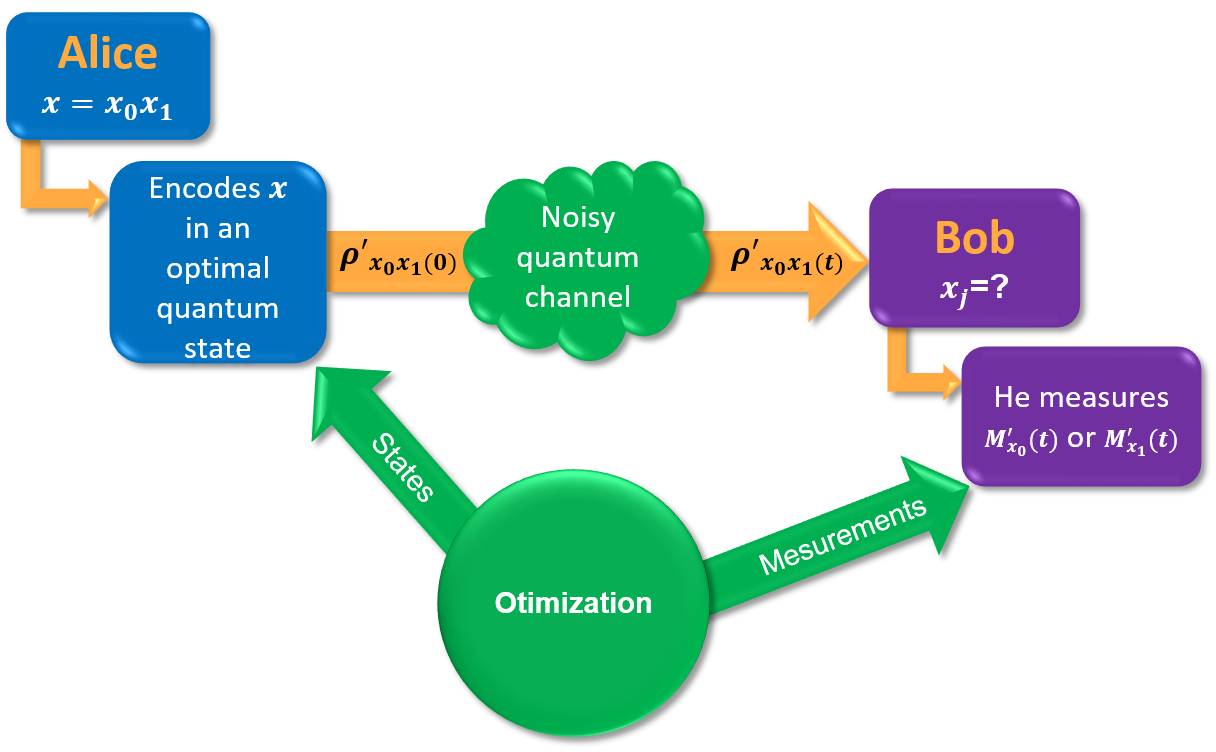}
    \caption{Schematics of the optimized noisy QRAC protocol.}
    \label{fig:quantico_noise_opt}
\end{figure}

The best strategy can be found for a QRAC when we take into account that $P^Q(t)$ does not depend only on the type of noise but also depends on whether or not the encoding states and decoding measurements are well adjusted to mitigate the noise channel effects. Therefore, choosing a good encoding requires anticipating the transformations that the channel may realize on the input states, and, whenever possible, optimize for the states that are least affected by them. Likewise, the decoding measurements should also reflect these channel-induced transformations. This new paradigm is illustrated in figure (\ref{fig:quantico_noise_opt}).


The optimal encoding and decoding are determined by maximizing $P^Q(t)$. We formulated this optimization problem in terms of two interdependent SDP sub problems:
\begin{equation}\label{sdp1}
\begin{cases}
\mathbf{max}\quad & P^Q(t) \\
\mathbf{s.t.:}\quad & \Tr{\rho'_{x_0x_1}(t)}=1,\\
& \rho'_{x_0x_1}(t)\succcurlyeq 0,
\end{cases}
\end{equation}
 and
\begin{equation}\label{sdp2}
\begin{cases}
\mathbf{max}\quad & P^Q(t) \\
\mathbf{s.t.:}\quad & \Tr{M'_{x_0}(t)}=1,\\
                    & \Tr{M'_{x_1}(t)}=1,\\
                    & M'_{x_0}(t)\succcurlyeq 0,\\
                    & M'_{x_1}(t)\succcurlyeq 0,
\end{cases}
\end{equation}
where $P^Q(t)$ is given by equation~\ref{pqt}. The optimization was conducted using the see-saw method \cite{seesaw}. In order to find the first iteration for the state $\rho'_{x_0x_1}(t)$ we run the sub problem describe in equation~\ref{sdp1} where the decoding measurements $M'_{x_0}(t)=\op{e_{x_0}}$ and $M'_{x_1}(t)=\op{f_{x_1}}$ are fixed. Next, we fixed the encoding, state $\rho'_{x_0x_1}(t)$ found in the previous iteration and optimize the decoding step, $M'_{x_0}(t)$ and  $M'_{x_1}(t)$ (equation~\ref{sdp2}). This process is repeated until the value of $P^Q(t)$ ceases to improve. We use Python 3.9 with the libraries Qutip for quantum mechanics operations and Picos for the SDP optimization using cvxopt solver.

\begin{figure}[H]
    \centering
    \includegraphics[scale=0.74]{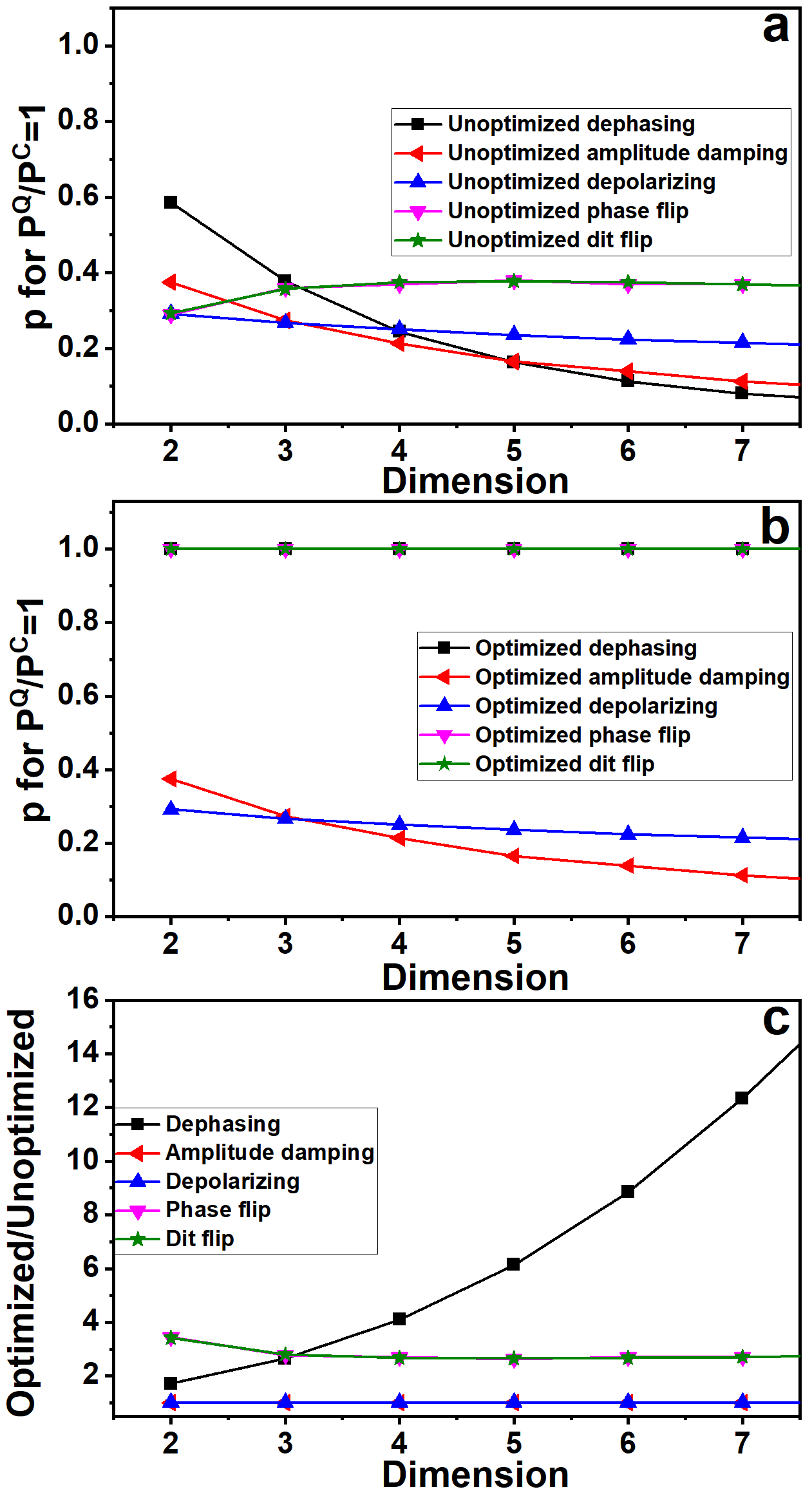}
    \caption{Graph \textbf{a}:  shows the values of $p$, where $p=1-\exp{\Gamma t}$, for which $P^Q/P^C=1$ as function of the dimension for the five noisy channels when QRAC is not optimized. Graph \textbf{b}: shows the values of $p$ for which $P^Q/P^C=1$ as function of the dimension for the five noisy channels when QRAC is optimized. Graph \textbf{c}: shows the ratio of the optimized to non-optimized values of $p$ as a function of the dimension, which expresses the gain achieved by the optimization.}
    \label{fig:graph0}
\end{figure}

In the graph \textbf{a} of figure (\ref{fig:graph0}) we show the value of $p$, where $p=1-\exp{-\Gamma t}$, when the threshold $P^Q=P^C$ was reached as a function of the dimension $d$, for the non-optimized QRAC, and in the graph \textbf{b} we show the same for the optimized QRAC. A comparison of the graphs \textbf{a} and \textbf{b} shows that the optimization was able to improve the performance of the QRAC for the dephasing channel, the dit flip channel, and the phase flip channel, where we saw that 
$$P^Q\geq P^C$$ for any time $t$ or dimension $d$. However, no improvement was found for the depolarizing and amplitude damping channels. In the graph \textbf{c} we show the ratio between the optimized and non-optimized values of $p$ (for $P^Q/P^C=1$) as a function of the dimension, which represents a measure of the improvement achieved by the optimization for each channel. We can see, for instance, that this improvement was overall greater for the dephasing channel but was also considerable for the dit flip and phase flip channels.

\begin{figure}[H]
    \centering
    \includegraphics[scale=0.78]{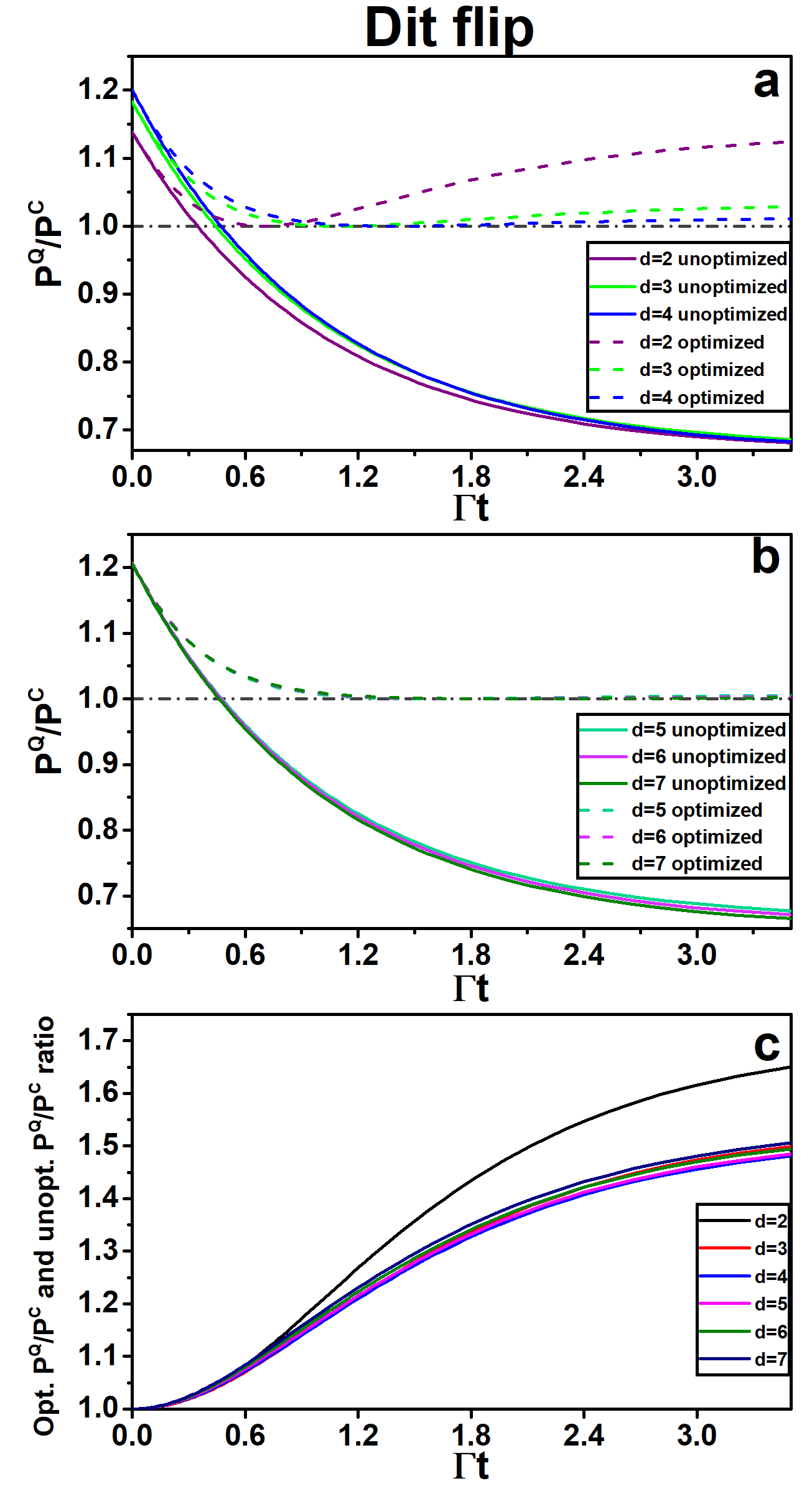}
    \caption{Graph \textbf{a} and \textbf{b}: The continuous lines show the \textit{non-optimized} $P^Q/P^C$ ratio as function of time, and the dashed lines represent this ratio after QRAC was optimized. Graph \textbf{c}: The optimized $P^Q/P^C$ ratio over the the non-optimized $P^Q/P^C$ one \textit{versus} time, in other words, it shows the gain achieved by the optimization.}
    \label{fig:graph1}
\end{figure}

\begin{figure}[H]
    \centering
    \includegraphics[scale=0.78]{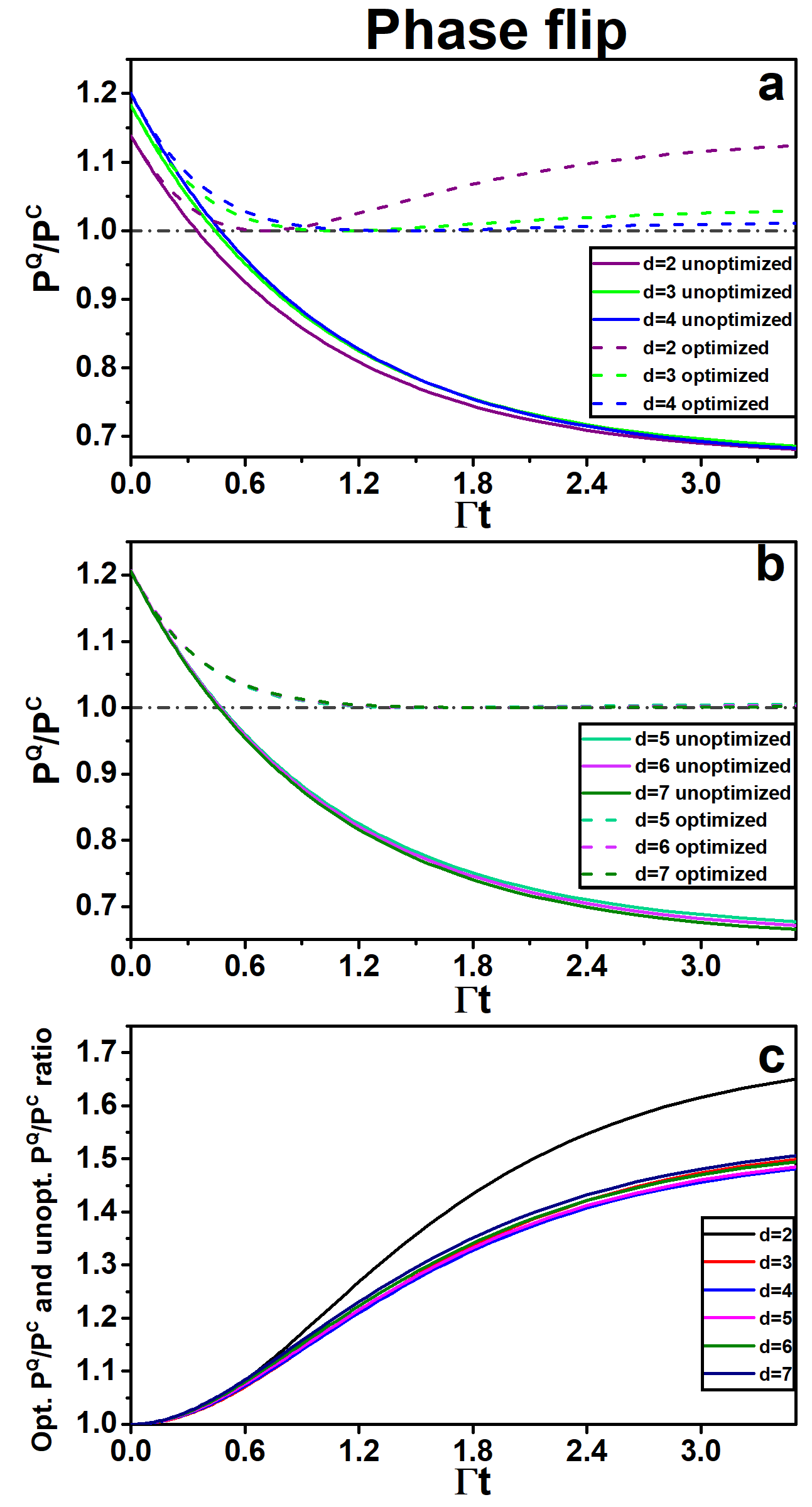}
    \caption{Graph \textbf{a} and \textbf{b}: The continuous lines show the \textit{non-optimized} $P^Q/P^C$ ratio as function of time, and the dashed lines represent this ratio after QRAC was optimized. Graph \textbf{c}: The optimized $P^Q/P^C$ ratio over the the unoptimized $P^Q/P^C$ one \textit{versus} time, in other words, it shows the gain achieved by the optimization.}
    \label{fig:graph2}
\end{figure}

\begin{figure}[H]
    \begin{center}
    \includegraphics[scale=0.31]{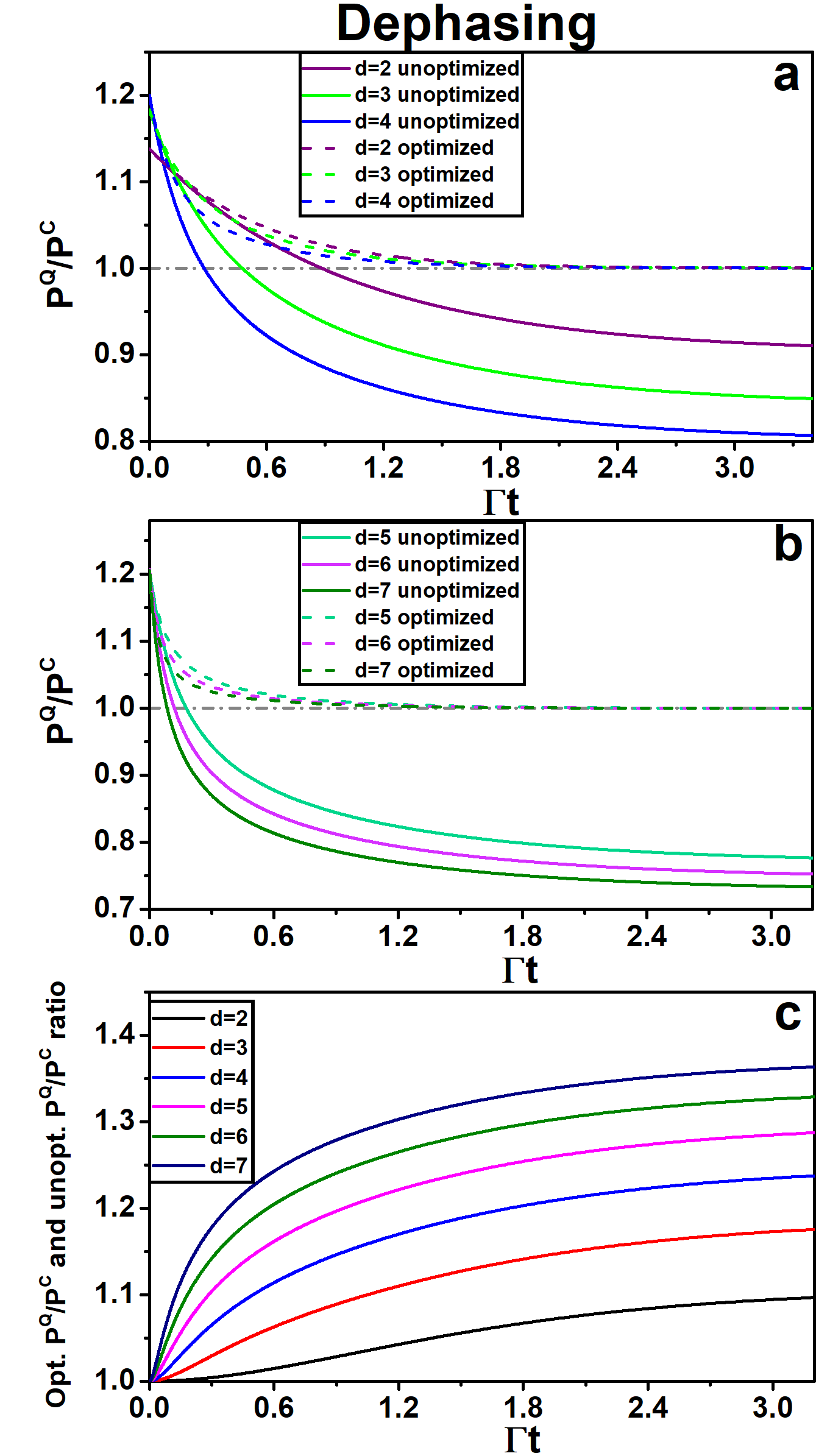}
    \caption{Graph \textbf{a} and \textbf{b}: The continuous lines show the \textit{non-optimized} $P^Q/P^C$ ratio as function of time, and the dashed lines represent this ratio after QRAC was optimized. Graph \textbf{c}: The optimized $P^Q/P^C$ ratio over the the unoptimized $P^Q/P^C$ one \textit{versus} time, in other words, it shows the gain achieved by the optimization.}
    \label{fig:graph3}
    \end{center}
\end{figure}

In the graphs \textbf{a} and \textbf{b} of figures (\ref{fig:graph1}), (\ref{fig:graph2}), and (\ref{fig:graph3}) the continuous lines represent the evolution of the non-optimized $P^Q/P^C$ ratio as a function of $\Gamma t$ while the dashed lines represent the optimized case for the dit flip, phase flip, and dephasing channels, respectively. A comparison of the dashed and continuous lines shows that the optimization substantially improves the values of $P^Q/P^C$ overall. It provides an advantage even for lower values of $\Gamma t$, and, for greater values, it guarantees that the performance of the QRAC is greater or equal to that of the CRAC. Particularly, for the dit flip and phase flip channel, it is worth noting that, for $d=2$ and $t\to \infty$ ($p=1$), the value of  $P^Q/P^C$ for the optimal QRAC approaches the value it had for $t=0$ (or the noiseless QRAC), something that does not happen for the non-optimal QRAC. This happens because at this limit the dit and the phase are completed flipped, and the optimal decoding measurements are the ones that take this fact into account. Therefore, keeping the measurements fixed in this situation is detrimental to the performance of the protocol. Moreover the optimization makes the QRAC more robust to high-dimensional dephasing noise, which tends to be greater the higher the dimension. In the graphs \textbf{c} of figures (\ref{fig:graph1}), (\ref{fig:graph2}), and (\ref{fig:graph3}) we display how many times the value of the optimal $P^Q/P^C$ ratio is greater than the non-optimal one as a function of $\Gamma t$, which represent the gain from optimization, that increases with $\Gamma t$. For the dephasing channel, it increases with the dimension, but for the dit flip and phase flip channels, it varies less with $d$ and has highest value happens for $d=2$.\\ 
\indent As we discussed in section (\ref{first}), the optimal QRAC strategy is based on MUB: the information is encoded in the superposition of states coming from two MUB, and it is decoded by measuring in either one of those bases, depending on which letter Bob is interested in. However, when optimizing the strategy in presence of a noisy quantum channel, the resulting encoding and decoding are not based exclusively on MUB anymore. Therefore, indiscriminately relying on MUB in such scenario can lead to sub-optimal encoding and decoding strategies. Among the maps that we considered, for the dit flip and phase flip the optimal decoding measurements are in MUB for all time $t$, however the encoding states are not necessarily based on two-state superposition of different MUB. For the dephasing, the encoding and the decoding change, but the computational bases is preserved as one of the measurement basis. In this work we focus on the QRAC protocols, however this symmetry loss might be true for other prepare-measure protocols. \\
 \indent Given the success of optimization for the dit flip, phase flip and dephasing channels, a natural question is why it failed for the depolarizing and amplitude damping channels. First, it should be emphasised that our approach only works when a better encoding states and better decoding measurements exist. The depolarizing channel transforms the qudit symmetrically (this can intuitively be seen in figure (\ref{fig:depo}) for $d=2$), and as consequence, the probabilities of success in any pair of encoding and decoding basis will decrease equally with time for any encoding states. Therefore, the encoding and decoding using  $B_e$ and $B_f$ already lead to the best $P^Q(t)$ one can possibly achieve for this channel. In the amplitude damping channel, the qudit drives it to the fundamental state which makes the probability of decoding some values of $x_0$ or $x_1$ increasingly small, and even null depending on the basis, as time passes. This happens, for instance, when we use the computational basis for encoding and decoding $x_0$. In this case, when $t\to \infty$ no information can be sent by this channel, independently of encoding and decoding strategy. Despite this, however, the fact that we fail to optimize the QRAC under amplitude damping channel for finite $t$ using the approach presented in this work does not necessarily imply it cannot be accomplished. Therefore, this matter should be subject of further investigation.\\
 \indent\section{Conclusion}
In this work we reviewed the concept of random access code in both its classical and quantum versions using qudits and the generalized quantum maps to describe well-known noisy channels for any discrete dimension. We built upon this work by incorporating the theory of open quantum systems to understand how a noisy channel affects the performance of QRAC and we showed how the see-saw method can be used to optimize the protocol even in the presence of losses. The method presented here can be useful for many other quantum communication protocols to improve their effectiveness without need of extra resources. This work opens new possibilities to investigate other methods to improve the use in quantum communication and quantum computation protocols that have mapped their errors. In this work we focused on a single-qudit encoding, but we aspect that the method can also be used more than one quantum system or different types of quantum information protocols that is not a prepare-measure protocol.

 \indent\section*{Acknowledgments}
R.A.S. acknowledges support from the Brazilian agency CAPES and CNPq. B.M. acknowledge partial support from the Brazilian National Institute of Science and Technology of Quantum Information (CNPq-INCT-IQ 465469/2014-0), CAPES/PrInt Process No. 88881.310346/2018-01 and CNPq (Grant No. 305165/2021-6).

\end{document}